\documentclass[prb,letterpaper,aps,floatfix,twocolumn]{revtex4-2}

\usepackage{graphicx}
\usepackage{amsmath,amssymb}
\usepackage{bm}

\usepackage[pdfstartview=FitH,breaklinks=true,bookmarks=true,colorlinks=true,anchorcolor=black,citecolor=red,filecolor=black,menucolor=black,urlcolor=blue,linkcolor=blue]{hyperref}

\usepackage{color}
\definecolor{gray}{rgb}{0.7,0.7,0.7}

\begin{document}

\title{Mitigating the sign problem by quantum computing}

\author{Kwai-Kong Ng}
\email{kkng@thu.edu.tw}
\affiliation{Department of Applied Physics, Tunghai University, Taichung 40704, Taiwan}
\author{Min-Fong Yang}
\email{mfyang@thu.edu.tw}
\affiliation{Department of Applied Physics, Tunghai University, Taichung 40704, Taiwan}

\date{\today}

\begin{abstract}
The notorious sign problem severely limits the applicability of quantum Monte Carlo (QMC) simulations, as statistical errors grow exponentially with system size and inverse temperature. A recent proposal of a quantum-computing stochastic series expansion (qc-SSE) method suggested that the problem could be avoided by introducing constant energy shifts into the Hamiltonian. Here we critically examine this framework and show that it does not strictly resolve the sign problem for Hamiltonians with non-commuting terms. Instead, it provides a practical mitigation strategy that suppresses the occurrence of negative weights. Using the antiferromagnetic anisotropic XY chain as a test case, we analyze the dependence of the average sign on system size, temperature, anisotropy, and shift parameters. An operator contraction method is introduced to improve efficiency. Our results demonstrate that moderate shifts optimally balance sign mitigation and statistical accuracy, while large shifts amplify errors, leaving the sign problem unresolved but alleviated.

\end{abstract}

\maketitle

\section{INTRODUCTION}

Quantum Monte Carlo (QMC) simulations are powerful and unbiased numerical tools for nonperturbative exploration of strongly correlated systems~\cite{QMC_book2016}, which are often analytically intractable. In such approaches, the partition function of an interacting system is written as a sum (or integral) over the exponentially large configurational space in a chosen basis. Through importance sampling over a small yet representative subset of the entire configuration space, QMC algorithms can provide an efficient method to investigate large-scale many-body systems in two or more dimensions with high accuracy.

Despite their advantages, QMC approaches are often hindered by the infamous ``sign problem''%
~\cite{Loh_etal1990,Hatano-Suzuki1992,Henelius-Sandvik2000,Troyer-Wiese2005} (for recent reviews, see Refs.~\cite{sign_review2021,sign_review2024}). This means that certain sampling weights in the QMC procedures are not positive definite, thereby precluding their interpretation as classical probabilities in the Monte Carlo framework. More importantly, configurations with negative weights cancel the contributions of those with positive weights, leading to an exponential growth of statistical errors with increasing system size and inverse temperature~\cite{Henelius-Sandvik2000,Troyer-Wiese2005}. The sign problem hence severely limits the applicability of QMC methods. From the outset, the quest to obviate or alleviate the sign problem has been a major research focus in the QMC community.

Extensive efforts over the years have revealed that the origins of the sign problem are multifaceted, lacking a single universal explanation. Consequently, the prospect of a generic solution appears unfeasible.
Nonetheless, numerous approaches have been proposed and explored to resolve or mitigate the problem in specific contexts~\cite{sign_review2024}. For the quantum spin systems considered here, the sign problem arises from positive off-diagonal elements in the Hamiltonian matrix, whose presence is basis-dependent~\cite{Hatano-Suzuki1992}.

In other words, the sign problem is closely related to the choice of basis, and numerous studies have shown that appropriate basis transformations can significantly alleviate or even eliminate it%
~\cite{Alet_etal2016,Honecker_etal2016,Ng-Yang2017,Stapmanns_etal2018,%
Wessel_etal2017,Wessel_etal2018,DEmidio_etal2020,Hangleiter_etal2020,%
Levy-Clark2021,Murota-Todo2025}.

While these methods may successfully prevent or mitigate the sign problem, they often rely on model-specific insights and physical intuition in selecting appropriate basis, and a systematic formulation remains challenging.

With the recent advent of quantum computation, substantial efforts have been devoted to developing quantum algorithms capable of accelerating specific computational tasks beyond the reach of classical methods (for recent reviews, see Refs.~\cite{Mazzola2024,Jiang_etal2025}). The ability of quantum computers to represent superpositions of states offers a promising avenue for overcoming limitations inherent in conventional approaches.

In this context, Tan \emph{et al.}~\cite{Tan_etal2022} recently proposed a novel implementation of the stochastic series expansion (SSE) QMC algorithm on a quantum computer, highlighting its significant advantages over classical counterparts. By leveraging quantum superposition, this quantum computing SSE (qc-SSE) approach enables efficient evaluation of matrix elements, thereby dispensing with the no-branching requirement that typically constrains classical implementations. Moreover, the choice of basis states is no longer restricted, provided they can be efficiently prepared on a quantum computer.
Building on these computational conveniences, Ref.~\cite{Tan_etal2022} proposes a general framework aimed at resolving the sign problem via quantum computing. They show that configuration weights can be rendered non-negative by adding a sufficiently large constant $M$ to each term in the Hamiltonian. Although this modification may violate the no-branching condition, it poses no obstacle to the evaluation of configuration weights within quantum algorithms. This conceptual shift could offer a practical resolution to the sign problem and broaden the applicability of the algorithm to a wider class of physical systems.
To validate their algorithm, the authors simulated a one-dimensional Ising spin chain Hamiltonian with $N=3$, 4, and 5 sites. The basis states were constructed as product states of Pauli $Z$ eigenstates, transformed via non-Clifford $T$ gates, resulting in quantum configurations that are classically intractable. In all cases examined, the average energy computed using this qc-SSE approach was found to converge to the exact finite-temperature energy obtained via exact diagonalization, thereby demonstrating the efficacy of the algorithm.

In the present work, we further examine the general validity of the qc-SSE approach in addressing the sign problem. We find that, for generic cases where the Hamiltonian contains non-commuting terms, the proposed framework does not resolve the sign problem in a strict sense.
According to Ref.~\cite{Tan_etal2022}, non-negative configuration weights can be guaranteed by choosing a shift constant $M=2n_\mathrm{c}$, where $n_\mathrm{c}$ is the cutoff in the expansion order. However, as explained in Sec.~\ref{sec:qc-SSE}, this setting introduces an inconsistency: the maximum operator string length during the simulation will exceed the predefined cutoff $n_\mathrm{c}$ that significant contributions beyond the cutoff are excluded from the sampling. As a result, this procedure will yield incorrect results.

Although the qc-SSE approach does not eliminate the sign problem entirely, it can still serve as a generic mitigation strategy by significantly reducing the occurrence of negative configuration weights. For illustration, we consider an antiferromagnetic XY spin chain and investigate the optimal choice of the constant $M$ for alleviating the sign problem. Here we employ an operator contraction method to enhance computational efficiency (see Appendix A). Without this technique, the simulations are restricted to much smaller system sizes owing to the prohibitive cost of evaluating long operator strings. Our results show that increasing $M$ significantly improves the average sign $\langle\textrm{sgn}\rangle$, which serves as a measure of the severity of the sign problem. However, this improvement comes at the cost of increased statistical errors in energy measurements, due to the increase in length of operator strings. Our findings suggest a practical trade-off, with $M=1$ offering a good balance between mitigating the sign problem and maintaining statistical accuracy. It is important to emphasize that the sign problem persists under this qc-SSE approach, particularly at low temperatures and in larger systems. Its severity also depends sensitively on system parameters, such as anisotropy. Consequently, the sign problem remains a fundamental obstacle to accurate and efficient simulation of quantum many-body systems.

The remainder of this paper is organized as follows.
The conventional SSE QMC method and the sign problem are briefly reviewed in Sec.~\ref{sec:SSE}.
We outline the framework of the qc-SSE in Sec.~\ref{sec:qc-SSE}, where its failure in resolving the sign problem is explained.
In Sec.~\ref{sec:results}, we introduce our model and then present our QMC results to show the $M$ dependence of the severity of the sign problem. The dependence of the average sign $\langle\textrm{sgn}\rangle$ on system sizes, temperature, and spin anisotropy are examined.
We conclude our paper in Sec.~\ref{sec:conclusion}.
The employed operator contraction method is described in Appendix A.

\section{SSE QMC and sign problem}\label{sec:SSE}

To make the presentation self-contained, we begin by introducing some basic concepts of the SSE method.
The SSE method is a widely adopted finite-temperature QMC algorithm for simulating quantum many-body systems~\cite{SSEreview_1,SSEreview_2}. It operates by stochastically sampling matrix elements from the Taylor series expansion of the density matrix in a suitably chosen basis. A significant advantage of SSE is its numerical exactness: the method introduces no systematic error from truncating the series, and its accuracy is limited solely by statistical uncertainties inherent in the sampling process.

For quantum systems with Hamiltonian $H$ in thermal equilibrium at inverse temperature $\beta$, the partition function is written as
\begin{equation}
Z=\mathrm{Tr}\left(e^{-\beta H}\right)=\sum_{\alpha} \langle\alpha|e^{-\beta H}|\alpha\rangle \;,
\end{equation}
where $\{|\alpha\rangle\}$ is some complete set of basis vectors. Defining $H\equiv-\sum_{b} H_b$ and employing the Taylor series expansion of the density matrix $e^{-\beta H}$, we have
\begin{equation}
Z=\sum_{n=0}^{\infty} \sum_{\{b_i\}} \sum_{\alpha} \frac{\beta^n}{n!} \langle\alpha|H_{b_n}\cdots H_{b_1} |\alpha\rangle \;.
\end{equation}
In practical calculations, the expansion power $n$ is always truncated at some sufficiently large value $n_\mathrm{c}$ to ensure computational feasibility. The set $\{b_i\}_{i=1}^n$ specifies an operator string of length $n$ emerging from taking the $n$-th power of the Hamiltonian, where each of $b_i$ indexes one of the terms in $H$.
Because the summation over all operator strings $\{b_i\}$ encompasses both $\langle\alpha|H_{b_n}\cdots H_{b_1} |\alpha\rangle$ and its complex conjugate $\langle\alpha|H_{b_1}\cdots H_{b_n} |\alpha\rangle$, only the real parts of the matrix elements have contributions.

The partition function is evaluated by generating a random walk in a configuration space $\{\mathcal{C}\}=\{(n, \{b_i\}, \alpha) \;\forall\, n, b_i, \alpha\}$ for different perturbation order $n$, index $\{b_i\}$, and state $|\alpha\rangle$, which are sampled under the assigned configuration weight
\begin{equation}
W(\mathcal{C})=\frac{\beta^n}{n!} \langle\alpha| H_{b_n}\cdots H_{b_1} |\alpha\rangle \;.
\label{eq:weight}
\end{equation}
When the no-branching condition, $H_{b_i} |\alpha\rangle \propto |\alpha'\rangle$, is satisfied in the computational basis, the simulation process avoids generating superpositions over basis states, thereby alleviating the exponential cost associated with evaluating and storing all diagonal and off-diagonal matrix elements in the operator string. In contrast, general superpositions of basis states are practically intractable to track on classical hardware, except for systems with limited size.

It is clear that the configuration weight $W(\mathcal{C})$ could be negative when some of the (off-diagonal) matrix elements of $H_{b_i}$ in the computational basis are negative. This leads to what is famously known as the sign problem of QMC.
In the presence of the sign problem, configuration weights cannot be treated as un-normalized probabilities as they should in Markov chain Monte Carlo simulations. In such cases, the usual expedient is to introduce a reference system with the configuration weights being the absolute values $|W_\mathcal{C}|$ of the original ones, and to add the sign into the sampling process of physical observables~\cite{Loh_etal1990,Hatano-Suzuki1992,Henelius-Sandvik2000,Troyer-Wiese2005}. This approach does not resolve the sign problem, but only casts it into a different form. By doing so, a thermal average over the configurations $\mathcal{C}$ of an observable $O$ is rewritten as
\begin{align}
\langle O\rangle
&=\frac{\sum_\mathcal{C} O_\mathcal{C}\;W_\mathcal{C}}{\sum_\mathcal{C} W_\mathcal{C}}
=\frac{\sum_\mathcal{C} O_\mathcal{C}\;\textrm{sgn}(W) |W_\mathcal{C}|/\sum_\mathcal{C} |W_\mathcal{C}|}{\sum_\mathcal{C} \textrm{sgn}(W) |W_\mathcal{C}|/\sum_\mathcal{C} |W_\mathcal{C}|} \nonumber \\
&= \frac{\langle O\;\textrm{sgn}(W)\rangle_{|W|}}{\langle \textrm{sgn}(W)\rangle_{|W|}} \;.
\end{align}
Here $\textrm{sgn}(W)=W_\mathcal{C}/|W_\mathcal{C}|$ denotes the sign of the original weight $W_\mathcal{C}$ and $\langle\cdots\rangle_{|W|}$ the thermal average with respect to the reference system with the configuration weights $|W_\mathcal{C}|$. Since the average sign $\langle\textrm{sgn}\rangle \equiv \langle\textrm{sgn}(W)\rangle_{|W|}$ enters the denominator of the estimator for expectation values, its magnitude critically affects the efficiency of the simulation. A small average sign amplifies statistical fluctuations, thereby increasing the runtime needed to achieve reliable results with controlled error bars. For models free of the sign problem, all configuration weights are positive, yielding $\langle\textrm{sgn}\rangle=1$, and the expression for $\langle O\rangle$ naturally reduces to its original form. In contrast, for sign-problematic systems where $\langle\textrm{sgn}\rangle\approx0$, the resulting error bars become prohibitively large, necessitating exponentially long simulation times to achieve accurate estimates of observables. Consequently, $\langle\textrm{sgn}\rangle$ is widely regarded as a key figure of merit for quantifying the severity of the sign problem.

In short, a crucial metric for quantifying the severity of the sign problem is the average sign $\langle\textrm{sgn}\rangle$. It can be expressed as the ratio between the partition function of the original system, $Z=\sum_\mathcal{C} W_\mathcal{C}$, and that of a reference system with absolute weights, $Z'=\sum_\mathcal{C} |W_\mathcal{C}|$, where $Z'\geq Z$ by construction. Typically, this average sign decays exponentially with increasing system size $N$ and inverse temperature $\beta$, following the form~\cite{Troyer-Wiese2005,sign_review2021,sign_review2024}
\begin{equation}\label{eq:avg_sign}
\langle\textrm{sgn}\rangle=\frac{Z}{Z'}=\exp(-N\beta\Delta f) \;,
\end{equation}
where $\Delta f\geq0$ denotes the difference in free energy densities between $Z$ and $Z'$. This exponential decay necessitates a correspondingly exponential increase in the number of samples to maintain statistical accuracy and control variance in observable estimates.

\section{quantum computing SSE method}\label{sec:qc-SSE}

In what follows, we restrict our analysis to spin-1/2 systems of size $N$, whose Hamiltonian can be written as
\begin{equation}
H=-\sum_{b} h_b\;\mathcal{O}_b
\end{equation}
with $\mathcal{O}_b$ being tensor products of Pauli matrices.

On quantum computers, superpositions of states are represented natively, enabling efficient evaluation of matrix elements in arbitrary basis. Consequently, the no-branching requirement is no longer necessary on the quantum device. To avoid negative weights during sampling, Tan \emph{et al.}~\cite{Tan_etal2022} propose adding sufficiently large constants
$M$ to each term in the Hamiltonian. They subsequently introduce a quantum circuit that efficiently evaluates the corresponding matrix elements, while the configuration space is sampled classically.

To be specific, the Hamiltonian is shifted by
\begin{equation}
H_b = h_b\,\mathcal{O}_b \rightarrow H_b = |h_b| \left[M+\mathrm{sgn}(h_b)\,\mathcal{O}_b\right] \;.
\label{eq:shift_M}
\end{equation}
Due to the violation of the no-branching requirement, conventional QMC methods implemented on a classical computer fail to handle this formulation. Specifically, evaluating the operator strings in Eq.~\eqref{eq:weight} involves successive applications of $H_{b_i}$, each of which generates a superposition of states, resulting in an exponential growth in the number of terms. In contrast, when executed on a quantum computer, the configuration weight at each Monte Carlo step can be evaluated directly, and the no-branching restriction plays no role in the calculations.

To evaluate the configuration weights for the shifted operators, one can introduce an ancilla qubit prepared in the state
\begin{equation}
|\phi_i\rangle = \sqrt{M/(M+1)}|0_i\rangle + \sqrt{1/(M+1)}|1_i\rangle
\end{equation}
for each operator in the operator string. As a result, the $i$-th normalized operator $H_{b_i}/|h_{b_i}|$ is promoted to a controlled unitary operation acting on the system register, conditioned on the corresponding ancilla qubit:
\begin{align}
&U_{b_i} |\alpha\rangle |0_i\rangle = |\alpha\rangle |0_i\rangle \\
&U_{b_i} |\alpha\rangle |1_i\rangle = \mathrm{sgn}(h_{b_i})\,\mathcal{O}_{b_i} |\alpha\rangle |1_i\rangle \;.
\end{align}
Therefore, for the state $|\Psi_\mathcal{C}\rangle \equiv |\alpha\rangle |\phi_1\rangle \cdots |\phi_n\rangle$ on a $(N + n)$-qubit circuit, we have
\begin{equation}
\langle\Psi_\mathcal{C}|\, U_{b_n} \cdots U_{b_1} |\Psi_\mathcal{C}\rangle
= \frac{\langle\alpha| H_{b_n}\cdots H_{b_1} |\alpha\rangle}
{(M+1)^n |h_{b_n} \cdots h_{b_1}|}   \;,
\label{eq:mat_element}
\end{equation}
and the configuration weight for the shifted Hamiltonian thus becomes
\begin{equation}
W(\mathcal{C})=\frac{\beta^n}{n!} (M+1)^n |h_{b_n}\cdots h_{b_1}|
\langle\Psi_\mathcal{C}|\, U_{b_n} \cdots U_{b_1} |\Psi_\mathcal{C}\rangle \;.
\label{eq:new_weight}
\end{equation}

In SSE, with the configurations sampled according to their weights in the partition function, the energy of the system can be efficiently evaluated using the expression~\cite{SSEreview_1,SSEreview_2}
\begin{equation}
E=-\frac{\left\langle n\right\rangle}{\beta} + M \sum_{b} |h_b| \;. \label{eq:energySSE}
\end{equation}
Here $\left\langle n\right\rangle$ is the average length of operator string per Metropolis loop. Note that the last term in Eq.~\eqref{eq:energySSE} is due to adding a constant $M$ to each term of the Hamiltonian in Eq.~\eqref{eq:shift_M}.

The validity of their proposal is demonstrated for the case of one-dimensional Ising spin chain with limited sizes of $N=3$, 4, 5. The basis states were constructed as product states of Pauli $Z$ eigenstates, transformed via non-Clifford $T$ gates, resulting in quantum configurations that are classically intractable. Notice that, in this case, all terms in the Hamiltonian are mutually commuting and the sign problem can be avoided simply by taking $M=1$.

However, in generic cases where the Hamiltonians contain non-commuting terms (say, the anisotropic XY model discussed in the next section), setting $M=1$ is insufficient to eliminate the sign problem. As shown in the Supplementary Information of Ref.~\cite{Tan_etal2022}, one must instead choose $M=2n_\mathrm{c}$, with $n_\mathrm{c}$ denoting the cutoff in the expansion order, to ensure that the weights in Eq.~\eqref{eq:new_weight} are non-negative.

However, we stress that, even though the configuration weights can be rendered non-negative through this choice of $M$, the proposed framework fails to yield correct results. According to Eq.~\eqref{eq:energySSE}, the average value of $\langle n\rangle$ scales linearly with the adjustable shift constant $M$ for a fixed energy $E$ at given system parameters and temperature. Consequently, choosing $M=2n_\mathrm{c}$ may result in $\langle n\rangle$ significantly exceeding the cutoff $n_\mathrm{c}$, which contradicts the initial assumption underlying the truncation. That is, with a fixed cutoff $n_\mathrm{c}$ and setting $M=2n_\mathrm{c}$ to eliminate negative weights, results subject to significant truncation error are inevitable, as some important configurations with $n>n_\mathrm{c}$ are excluded from sampling. Alternatively, if the Taylor series expansion is left untruncated and an arbitrary finite value of $M$ is chosen, the non-negativity of weights for configurations with $n>>M$ is no longer guaranteed. In other words, the qc-SSE formalism does not resolve the sign problem in a strict sense, but rather recasts it into a different guise.

Nevertheless, as shown in the next section, this qc-SSE approach can be employed to mitigate the severity of the sign problem, as the introduction of a constant $M$ into the system Hamiltonian significantly suppresses the occurrence of negative configuration weights.

\section{mitigating sign problem by qc-SSE}\label{sec:results}

In this section, we examine the performance of the qc-SSE approach in mitigating the sign problem by applying it to the antiferromagnetic anisotropic XY (XZ) spin chain of length $N$ with periodic boundary conditions. The system is governed by the Hamiltonian
\begin{equation}
H=\sum_{i}{Z_i Z_{i+1}}+\Delta\sum_{i}{X_i X_{i+1}} \;,
\label{eq:Hami}
\end{equation}
where $Z_i$ and $X_i$ denote the Pauli matrices acting on site $i$, and the parameter $\Delta$ controls the anisotropy between the $ZZ$ and $XX$ interaction terms. It is important to note that operators in the first and second terms of $H$ do not commute in general, which complicates the elimination of the sign problem. In particular, setting $M=1$, as employed in Ref.~\cite{Tan_etal2022}, is insufficient to fully resolve the issue in this model. This system typically suffers from a negative sign problem, except when the chain length $N$ is even. Under such conditions, a $\pi$ rotation about the $y$ axis applied to one sublattice transforms the system into a ferromagnetic form, ensuring non-negative configuration weights within the standard SSE framework.
In the following, we consider only the case of $\Delta\le1$. The conclusions for the $\Delta>1$ case can be achieved simply by interchanging $Z_i\leftrightarrow X_i$ with energy rescaling.

Since our purpose is to mitigate rather than strictly resolve the sign problem, the requirement $M=2n_\mathrm{c}$ can be relaxed. Accordingly, we introduce distinct constants for different terms in the Hamiltonian, i.e.,
\begin{align}
H=&-\sum_{i}{\left(M_z - Z_i Z_{i+1} \right)} - \Delta\sum_{i}{\left(M_x - X_i X_{i+1}\right)} \nonumber \\
& + \left(M_z + \Delta M_x\right)N \;,
\label{eq:H_Mx_Mz}
\end{align}
and explore their optimal values to balance sign-problem suppression with computational efficiency. We note that $M_x$ and $M_z$ serve as tunable parameters and may take arbitrary positive real values.

In the qc-SSE approach, we sample the partition function by computing configuration weights using quantum circuits that represent the operator strings. This computation becomes increasingly demanding as the number of qubits grows, particularly when simulated on classical hardware. As discussed in the previous section, evaluating the configuration weight requires introducing an ancilla qubit for each term in the operator string. Consequently, the total number of qubits in the quantum circuit equals the operator string length $n$ plus the system size $N$, i.e., $N+n$ qubits are needed to implement the computation. When the operator string becomes long, the associated computational cost can become prohibitively large, especially in classical emulation scenarios. To improve computational tractability, we introduce an operator contraction method (see Appendix A) that compresses the operator string length without compromising accuracy. This approach markedly boosts simulation speed and allows us to investigate systems of size up to $N=7$ that would otherwise be computationally prohibitive.

In the following, we begin by identifying the preferred values of the constants $M_x$ and $M_z$ in Eq.~\eqref{eq:H_Mx_Mz}. For systematic comparison, all reported results are based on averages over 20,000 Monte Carlo sweeps. Each sweep comprises three components: updates of the basis state $|\alpha\rangle$, updates of the operator string $\{b_i\}$ at fixed length, and insertion/removal updates of the operator string, corresponding to $n\rightarrow n\pm1$.

\subsection{Optimizing shift constants $M_x$ and $M_z$}

\begin{figure}[t]
\includegraphics[width=1\linewidth]{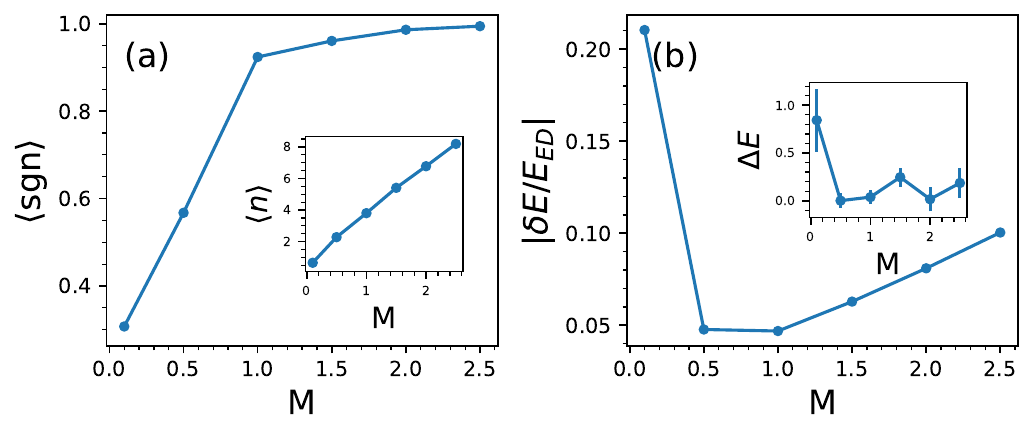}
\includegraphics[width=1\linewidth]{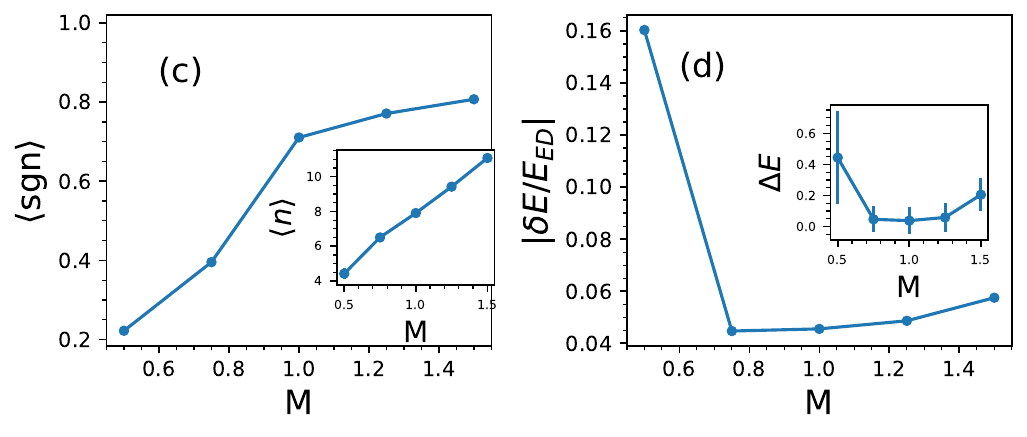}
\caption{ (a) Average sign $\langle \textrm{sgn} \rangle$ and (b) absolute percentage error $|\delta E / E_\mathrm{ED}|$ as functions of $M = M_x = M_z$ for system size $N=3$ and temperature $T=2$. The inset of (a) shows the average operator string length $\langle n\rangle$, while the inset of (b) shows the absolute energy difference  $\Delta E = |E - E_\mathrm{ED}|$. Panels (c) and (d) present the corresponding results at temperature $T=1$.}
\label{fig:M_dep}
\end{figure}

As in Ref.~\cite{Tan_etal2022}, our simulations employ basis states formed from product states of Pauli $Z$ eigenstates, modified by a Hadamard gate and non-Clifford $T$ gate transformations. In this representation, both components of the Hamiltonian in Eq.~\eqref{eq:H_Mx_Mz} contribute off-diagonal elements, which can lead to negative configuration weights irrespective of the system size $N$. As discussed above, introducing the constants $M_x$ and $M_z$ can mitigate the sign problem, though it does not eliminate it entirely. Here, we systematically investigate the impact of these parameters on the average sign and the precision of total energy measurements in the antiferromagnetic XY spin chain.

We begin by setting $M_x=M_z=M$ to investigate how the average sign $\langle\textrm{sgn}\rangle$ depends on $M$ for the isotropic ($\Delta=1$) spin chain of size $N=3$ at temperature $T=2$. Using the qc-SSE algorithm, our results in Fig.~\ref{fig:M_dep}(a) indicate that $\langle\textrm{sgn}\rangle$ increases with larger values of $M$, reflecting reduced severity of the sign problem. Starting from $\langle\textrm{sgn}\rangle=0.31$ at $M=0.1$, it rises almost linearly to 0.92 at $M=1.0$, after which the increase becomes more gradual, approaching nearly 1 for $M=2.5$. The severity of the sign problem at small $M<1$ can be attributed to the spectral properties of the operators $X_iX_{i+1}$ and $Z_iZ_{i+1}$, whose largest eigenvalue is 1. Selecting $M$ below this bound leads to frequent occurrences of negative weights during sampling, thereby sustaining a pronounced sign problem.

To ensure that all possible configurations are accessible during sampling, we do not truncate the operator strings in our simulations. Empirically, the required cutoff $n_\mathrm{c}$ is typically several times larger than the average length $\langle n\rangle$. From our data for $\langle n\rangle$ presented in the inset of Fig.~\ref{fig:M_dep}(a), it follows that $n_\mathrm{c}$ must significantly exceed the value of $M$. Therefore, as discussed above, the prescription $M=2n_\mathrm{c}$ will result in misleading outcomes and fails to capture the correct behavior of the system.

Given the observed monotonic improvement of the average sign with increasing $M$, it appears that larger values of $M$ are advantageous within the qc-SSE framework. Nonetheless, a different perspective arises when considering the absolute percentage error $|\delta E/E_\mathrm{ED}|$, with $\delta E$ representing the standard deviation of the simulated energy [as defined in Eq.~\eqref{eq:energySSE}] and $E_\mathrm{ED}$ denoting the exact diagonalization benchmark. Fig.~\ref{fig:M_dep}(b) reveals that for $M>1$, the absolute percentage error $|\delta E / E_\mathrm{ED}|$ increases with $M$. This behavior arises from $\delta E \propto \delta \langle n \rangle$ [as indicated by Eq.~\eqref{eq:energySSE}], combined with the well-known result $\delta \langle n \rangle \propto \sqrt{\langle n \rangle}$~\cite{SSEreview_1,SSEreview_2}. Due to the monotonic dependence of $\langle n \rangle$ on $M$ as seen from the inset of Fig.\ref{fig:M_dep}(a), larger values of $M$ induce greater fluctuations in energy estimates, thereby reducing accuracy despite improved average sign.
%
%
On the other hand, a small value of average sign $\langle\textrm{sgn}\rangle$ for $M<1$ also leads to large statistical errors as shown in Fig.~\ref{fig:M_dep}(b).
This trade-off is further highlighted by the energy difference $\Delta E = |E - E_\mathrm{ED}|$ in the insets of Fig.~\ref{fig:M_dep}(b): at $M=1$, the estimated energy agrees well with exact diagonalization with small uncertainty, whereas for $M$ either larger than or much smaller than 1, growing statistical errors lead to noticeable discrepancies.

Similar qualitative behavior is observed at the lower temperature $T=1$.  However, as shown in Figs.~\ref{fig:M_dep}(c) and (d), the average sign remains small and may not converge to unity even for large values of $M$. This indicates that the sign problem becomes significantly more severe compared to the case at $T=2$, and cannot be effectively mitigated by small $M$. As noted earlier, there is no guarantee that the sign problem can be eliminated when $M < 2n_\mathrm{c}$. In practice, choosing a large $M$ not only increases computational cost due to longer operator strings, which nearly double at $T=1$, but also amplifies statistical errors, rendering results unreliable. Nevertheless, setting $M$ close to 1 provides a practical compromise, sufficiently mitigating the sign problem while maintaining accurate computation of observables.

\begin{figure}[t]
\includegraphics[width=1\linewidth]{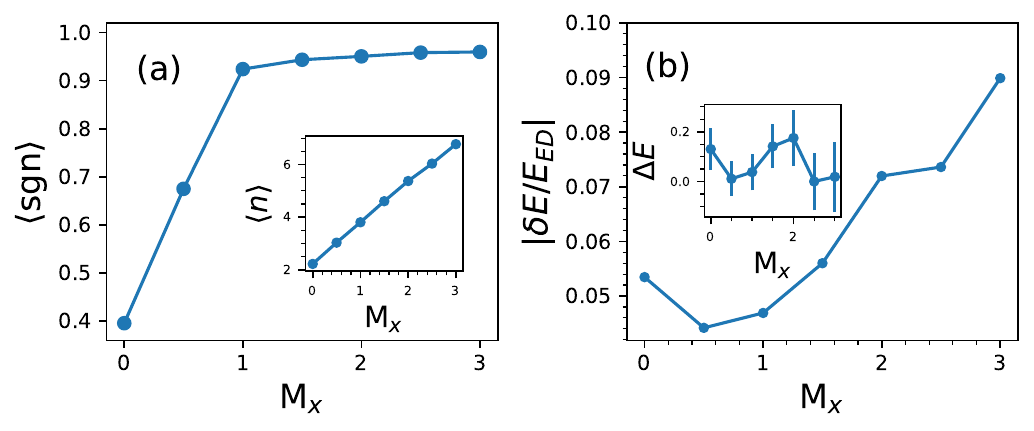}
\caption{(a) Average sign $\langle \textrm{sgn} \rangle$ and (b) absolute percentage error $|\delta E / E_\mathrm{ED}|$ as functions of $M_x$ with $M_z$ fixed at 1 for system size $N=3$ and temperature $T=2$. The inset of (a) shows the average operator string length  $\langle n\rangle$, while the inset of (b) shows the absolute energy difference $\Delta E = |E - E_\mathrm{ED}|$. }
\label{fig:M_dep_T_1}
\end{figure}

One may wonder whether relaxing the condition $M_x=M_z$ could yield better performance. To explore this possibility, we fix $M_z=1$ and vary $M_x$ at temperature $T=2$, with the corresponding results presented in Fig.~\ref{fig:M_dep_T_1}. The observed trends in average sign and absolute percentage error mirror the previous case: $\langle \textrm{sgn}\rangle$ improves significantly as $M_x$ increases, while $|\delta E / E_\mathrm{ED}|$ grows with $M_x$ when $M_x>1$. These findings indicate that setting $M_x=M_z=1$ strikes an effective balance between sign-problem mitigation and energy accuracy, and is thus adopted as the default choice in subsequent simulations.

\subsection{Dependence on system sizes and temperatures}

\begin{figure}[t]
\includegraphics[width=1\linewidth]{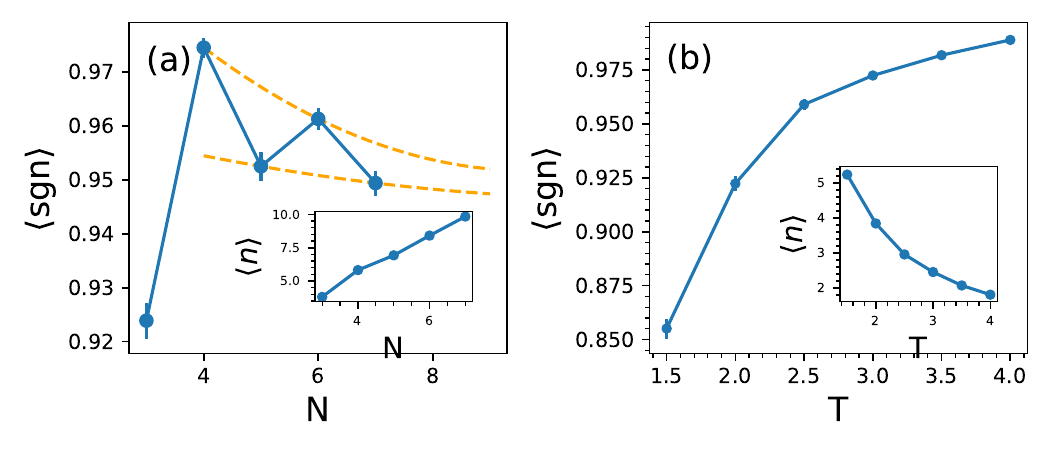}
\caption{(a) Size dependence of average sign $\langle \textrm{sgn} \rangle$ with fixed $T=2$; (b) temperature dependence of average sign $\langle \textrm{sgn} \rangle$ with fixed $N=3$. Here we take $M_x=M_z=1$.  The inset of (a) shows the average operator string length $\langle n\rangle_{|W|}$.}
\label{fig:size_dep}
\end{figure}

With the optimal choice $M_x=M_z=1$ established, we now examine how the performance of the qc-SSE approach varies with system size and temperature.

Figure~\ref{fig:size_dep}(a) presents our data on the system-size dependence of the average sign $\langle\textrm{sgn}\rangle$ at temperature $T=2$, for system sizes up to $N=7$. Notably, results for $N>4$ were made accessible within reasonable computation time through the use of the operator contraction technique detailed in the Appendix A.
As seen from our findings, the average sign for $N>3$ decreases with increasing system size $N$, accompanied by a pronounced even–odd oscillatory behavior.

The decay in $\langle\textrm{sgn}\rangle$, shown separately for both even and odd $N$, aligns with the exponential suppression predicted by Eq.~\eqref{eq:avg_sign}. The observed even–odd effect, wherein systems with odd $N$ exhibit smaller average signs than those with even $N$, can be attributed to geometric phases associated with the chordless cycles in the Hamiltonian~\cite{Hen2021}. This can be easily understood when working in the $z$-basis, where only the $X_iX_{i+1}$ terms act as off-diagonal operators and contribute negative matrix elements to the configuration weights.
To restore the original configuration $|\alpha\rangle$, certain operator strings generate cyclic paths in the configuration space, wherein the off-diagonal $X_iX_{i+1}$ bonds cover the chain an odd number of times. For odd $N$, these operator strings contain an odd number of $X_iX_{i+1}$ terms, resulting in negative contributions to the configuration weights in Eq.~\eqref{eq:weight}. This leads to sign cancellations and yield $\langle\textrm{sgn}\rangle<1$ for odd-length chains. In contrast, for even $N$, all contributing operator strings involve an even number of $X_iX_{i+1}$ terms, ensuring that the configuration weights are non-negative. Consequently, systems with even $N$ are sign-problem-free in the $z$-basis, yielding $\langle\textrm{sgn}\rangle=1$. This even-odd effect persists in the present basis choice, as clearly reflected in our numerical data. The occurrence of negative weights in even $N$ chains is less frequent, though not entirely eliminated, compared to odd $N$ chains.

Despite the overall decaying trend, the average sign remains close to unity, $\langle\textrm{sgn}\rangle\sim1$, within the range of available system sizes. This shows that the sign problem can be effectively mitigated by the qc-SSE approach for modest system sizes. Nevertheless, simulations of significantly larger systems may still pose substantial challenges due to the expected exponential decay of the average sign. Furthermore, since the implementation requires $N+n$ qubits, the computational cost remains demanding for large system sizes. This is evidenced by the increasing trend in the average operator string length $\langle n\rangle$ with system size $N$, as shown in the inset of Fig.~\ref{fig:size_dep}(a).
%

The temperature dependence of the average sign $\langle\textrm{sgn}\rangle$ for the $N=3$ spin chain is shown in Fig.~\ref{fig:size_dep}(b). As the temperature increases ($T=1/\beta$), we observe a corresponding rise in the average sign, in agreement with the theoretical prediction in Eq.~\eqref{eq:avg_sign}. Additionally, as shown in the inset, the average operator string length $\langle n\rangle$ decreases with increasing temperature. These observations suggest that the qc-SSE approach is effective and computationally efficient in the high-temperature regime.

\subsection{Anisotropy dependence}

\begin{figure}[t]
\includegraphics[width=0.8\linewidth]{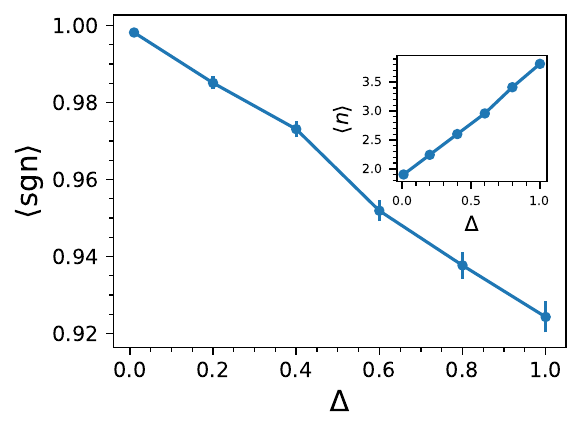}
\caption{Anisotropy dependence of average sign $\langle \textrm{sgn} \rangle$ for $N=3$ and $M_z=M_x=1$ at $T=2$.  The inset shows the average operator string length.}
\label{fig:delta_dep}
\end{figure}

Thus far, our analysis has focused on the isotropic limit, corresponding to the anisotropy parameter $\Delta=1$. We now investigate how the performance of the qc-SSE approach varies with $\Delta$, while maintaining the optimal parameter choice $M_x=M_z=1$ established earlier. Without loss of generality, we restrict our exploration to the case $\Delta\le1$, as the conclusions for $\Delta>1$ can be readily obtained by interchanging $Z_i\leftrightarrow X_i$ and applying an appropriate energy rescaling.

The anisotropy dependence of the average sign for the $N=3$ spin chain at temperature $T=2$ is shown in Fig.~\ref{fig:delta_dep}. We observe that the average sign increases monotonically and approaches unity as the anisotropy parameter $\Delta$ decreases toward zero, indicating a gradual alleviation of the sign problem with reduced anisotropy. Notably, the Ising limit with $\Delta=0$ has already been investigated in Ref.~\cite{Tan_etal2022} within the qc-SSE framework. In that study, the model was shown to be sign-problem-free even in a general basis. This arises from the fact that, in this limit, all terms in the Hamiltonian mutually commute, ensuring that the configuration weights become positive semidefinite in any basis when choosing $M_z=1$. Our results in this limit are consistent with their conclusions. Furthermore, this perspective offers a natural explanation for the observed gradual mitigation of the sign problem as the anisotropy parameter $\Delta$ decreases. Specifically, the sampling probability of the non-commuting terms $M_x-X_i X_{i+1}$ is suppressed with decreasing $\Delta$, thereby reducing the extent of destructive interference and improving the average sign.

The inset of Fig.~\ref{fig:delta_dep} reveals that the average operator-string length decreases as the anisotropy parameter $\Delta$ is reduced. This trend indicates that simulations become computationally less demanding in regimes far from the isotropic limit. These observations suggest that the qc-SSE approach is effective and computationally efficient in the weakly anisotropic case, thereby enabling access to lower-temperature regimes and/or larger system sizes that would otherwise be challenging.

\section{conclusions}\label{sec:conclusion}

In the present study, we investigate the capability and performance of the qc-SSE approach in mitigating the sign problem encountered in simulations of the antiferromagnetic XY spin chain. To further improve computational efficiency, we introduce an operator string contraction method that reduces the effective string length and accelerates weight evaluation.

Contrary to the conclusion presented in Ref.~\cite{Tan_etal2022}, we show that the introduction of constant shifts $M_x$ and $M_z$ cannot fully eliminate the sign problem. Regardless of how large these constants are chosen, the problem persists, particularly at low temperatures and for large system sizes. As illustrated in Sec.~\ref{sec:results}, increasing $M_x$ and $M_z$ does lead to a notable improvement in the average sign $\langle\textrm{sgn}\rangle$, thereby alleviating the severity of the sign problem. However, this improvement comes at the cost of longer operator strings, which in turn amplify statistical errors in energy measurements. A practical compromise is achieved by setting $M_x=M_z=1$, a choice that effectively suppresses negative weights without substantially inflating sampling errors, thus balancing accuracy and efficiency.

We note that the present approach alleviates the sign problem more effectively than the standard classical SSE, as illustrated in Appendix B. In addition, the classical framework offers no general strategy for eliminating negative weights across arbitrary Hamiltonians. A traditional approach is to search for a basis in which the Hamiltonian remains non-branching and, ideally, exhibits a reduced sign problem%
~\cite{Alet_etal2016,Honecker_etal2016,Ng-Yang2017,Stapmanns_etal2018,%
Wessel_etal2017,Wessel_etal2018,DEmidio_etal2020,Hangleiter_etal2020,%
Levy-Clark2021,Murota-Todo2025}. However, changing the basis typically requires rewriting the directed-loop equations and modifying the loop-update algorithm, which becomes increasingly complicated when multiple spins are coupled in the new basis. In contrast, qc-SSE entails no substantial algorithmic modifications and can be applied directly to general Hamiltonians in any basis, thereby mitigating negative signs regardless of branching.

Although the qc-SSE framework can be employed to alleviate the sign problem for arbitrary Hamiltonians and in general bases, its circuit depth, as indicated by $\langle n\rangle$, is found to scale with the inverse temperature $\beta$ and system size $N$. As a result, contributions from increasingly long operator strings become significant at low temperatures and for large systems, which limits the practical applicability of the method. 

A promising direction for further improvement lies in optimizing the choice of basis states used to construct the quantum circuits. Since the severity of the sign problem is inherently basis dependent, variationally tuning the local basis—such as optimizing the orientation of single-spin or multi-spin basis states—may further suppress negative-weight contributions without introducing additional algorithmic complexity. Within the qc-SSE framework, such basis optimization can be implemented at the circuit level without modifying the Monte Carlo update structure.

\begin{acknowledgments}
The authors would like to thank Ching-Yu Huang for enlightening discussions. This research was supported by Grant No. NSTC 114-2112-M-029-001, NSTC 114-2112-M-029-005, and NSTC 114-2112-M-029-006 of the National Science and Technology Council of Taiwan.
\end{acknowledgments}


\appendix
\section{Operator contraction}

By introducing a constant shift to each term in the Hamiltonian in Eq.~\eqref{eq:Hami}, the expression can be rewritten as [i.e., Eq.~\eqref{eq:H_Mx_Mz}]
\begin{align}
H=&-\sum_{i}{\left(M_z - Z_i Z_{i+1} \right)} - \Delta\sum_{i}{\left(M_x - X_i X_{i+1}\right)} \nonumber \\
& + \left(M_z + \Delta M_x\right)N \;.
\end{align}

When $M_z =1$, by using the properties of Pauli matrices: $(Z_i)^2=1$, $(X_i)^2=1$, and $Z_i X_i=-X_i Z_i$, we have the following operator identities:
\begin{widetext}
\begin{align}
\left(1 - Z_i Z_{i+1}\right)\left(M_x - X_i X_{i+1}\right)
&=\left(M_x - X_i X_{i+1}\right)\left(1 - Z_i Z_{i+1}\right) \;, \\
\left(1 - Z_i Z_{i+1}\right)\left(1 - Z_i Z_{i+1}\right)
&=2 \left(1 - Z_i Z_{i+1}\right) \;, \\
\left(1 - Z_i Z_{i+1}\right) \left(M_x - X_{i-1} X_i\right)
\left(1 - Z_i Z_{i+1}\right)
&=2M_x \left(1 - Z_i Z_{i+1}\right) \;, \\
\left(1 - Z_i Z_{i+1}\right) \left(M_x - X_{i+1} X_{i+2}\right)
\left(1 - Z_i Z_{i+1}\right)
&=2M_x \left(1-Z_iZ_{i+1}\right) \;.
\end{align}
%

In short, eliminating a term of the form $\left(1 - Z_i Z_{i+1}\right)$ yields a factor of 2, while eliminating $\left(M_x - X_i X_{i+1}\right)$ contributes a factor of $2M_x$. When both $M_z$ and $M_x$ differ from unity, the same principle applies, provided the constants are allowed to vary during the contraction process. For example, the following contraction produces a factor of $(M_1+M_2)$ and modifies the constant in front of the $Z_i Z_{i+1}$ accordingly:

%
%
 \begin{equation}
\left(M_1 - Z_i Z_{i+1}\right)\left(M_2 - Z_i Z_{i+1}\right)
=(M_1+M_2)\left(\frac{M_1 M_2+1}{M_1+M_2}- Z_i Z_{i+1}\right) \;.
\end{equation}

By applying the operator contraction rules outlined above, we are able to compress the operator string length without compromising accuracy. This approach significantly accelerates the simulation process and enables the exploration of systems up to size $N=7$, which would otherwise be computationally prohibitive due to resource constraints.

\end{widetext}

\section{Comparison with classical SSE}

\begin{figure}[t]
\includegraphics[width=0.8\linewidth]{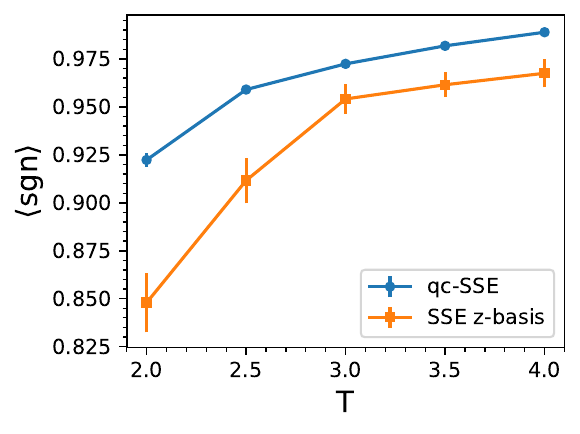}
\caption{Average sign $\langle \textrm{sgn} \rangle$ of classical SSE on z-basis and qc-SSE as a function of $T$ for $\Delta=1$ and $N=3$. }
\label{fig:cSSE_qcSSE}
\end{figure}

To demonstrate the advantage of using qc-SSE over the standard classical SSE, we compare the average sign $\langle \textrm{sgn} \rangle$ obtained from both methods as a function of temperature $T$ for the case of $\Delta=1$ and $N=3$.  For qc-SSE, we set $M_z=M_x=1$. By contrast, for the classical SSE on $z$-basis, $M_z=1$ but $M_x$ must be set to zero, since any finite $M_x$ introduces branching processes that the classical SSE cannot handle. As shown in Fig.~\ref{fig:cSSE_qcSSE}, the average sign produced by the proposed qc-SSE remains significantly higher than that of the classical SSE, highlighting the improved sign performance of the qc-SSE approach. It can be understood that introducing a finite $M_z$ and $M_x$ suppresses the negative contribution arising from the $Z_i Z_{i+1}$ and $X_i X_{i+1}$ operators, thereby allievating the sign problem.





\begin{thebibliography}{99}

\bibitem{QMC_book2016}
J. Gubernatis, N. Kawashima, and P. Werner, \emph{Quantum Monte Carlo Methods: Algorithms for Lattice Models} (Cambridge University Press, Cambridge, UK, 2016).


\bibitem{Loh_etal1990}
E. Y. Loh, J. E. Gubernatis, R. T. Scalettar, S. R. White, D. J. Scalapino, and R. L. Sugar, Sign problem in the numerical simulation of many-electron systems,
\href{https://doi.org/10.1103/PhysRevB.41.9301}
{Phys. Rev. B \textbf{41}, 9301 (1990)}.
%
\bibitem{Hatano-Suzuki1992}
N. Hatano, M. Suzuki, Representation basis in quantum Monte Carlo calculations and the negative-sign problem,
\href{https://doi.org/10.1016/0375-9601(92)91006-D}
{Phys. Lett. A \textbf{163}, 246 (1992)}.
%
\bibitem{Henelius-Sandvik2000}
P. Henelius and A. W. Sandvik, Sign problem in Monte Carlo simulations of frustrated quantum spin systems,
\href{https://doi.org/10.1103/PhysRevB.62.1102}
{Phys. Rev. B \textbf{62}, 1102 (2000)}.
%
\bibitem{Troyer-Wiese2005}
M. Troyer and U.-J. Wiese, Computational Complexity and Fundamental Limitations to Fermionic Quantum Monte Carlo Simulations,
\href{http://dx.doi.org/10.1103/PhysRevLett.94.170201}
{Phys. Rev. Lett. \textbf{94}, 170201 (2005)}.
%


\bibitem{sign_review2021}
D. Banerjee, Recent progress on cluster and meron algorithms for strongly correlated systems,
\href{https://doi.org/10.1007/s12648-021-02155-5}
{Indian J. Phys. \textbf{95}, 1669 (2021)}.
%
\bibitem{sign_review2024}
G. Pan and Z. Y. Meng, Sign problem in quantum Monte Carlo simulation,
\href{https://dx.doi.org/10.1016/B978-0-323-90800-9.00095-0}
{ECMP \textbf{1}, 879 (2024)}.


\bibitem{Alet_etal2016}
F. Alet, K. Damle, and S. Pujari, Sign-problem-free Monte Carlo simulation of certain frustrated quantum magnets,
\href{https://doi.org/10.1103/PhysRevLett.117.197203}
{Phys. Rev. Lett. \textbf{117}, 197203 (2016)}.
%
\bibitem{Honecker_etal2016}
A.~Honecker, S.~Wessel, R.~Kerkdyk, T.~Pruschke, F.~Mila, and B.~Normand, Thermodynamic properties of highly frustrated quantum spin ladders: influence of many-particle bound states,
\href{https://doi.org/10.1103/PhysRevB.93.054408}
{Phys. Rev. B \textbf{93}, 054408 (2016)}.
%
\bibitem{Ng-Yang2017}
K.-K. Ng and M.-F. Yang, Field-induced quantum phases in a frustrated spin-dimer model: a sign-problem-free quantum monte carlo study,
\href{https://doi.org/10.1103/PhysRevB.95.064431}
{Phys. Rev. B \textbf{95}, 064431 (2017)}.
%
\bibitem{Stapmanns_etal2018}
J.~Stapmanns, P.~Corboz, F.~Mila, A.~Honecker, B.~Normand, and S.~Wessel, Thermal critical points and quantum critical end point in the frustrated bilayer Heisenberg antiferromagnet,
\href{https://doi.org/10.1103/PhysRevLett.121.127201}
{Phys. Rev. Lett. \textbf{121}, 127201 (2018)}.


\bibitem{Wessel_etal2017}
S. Wessel, B.~Normand, F. Mila, and A. Honecker, Efficient quantum Monte Carlo simulations of highly frustrated magnets: the frustrated spin-1/2 ladder,
\href{https://10.21468/SciPostPhys.3.1.00}
{SciPost Phys. \textbf{3}, 005 (2017)}.
%
\bibitem{Wessel_etal2018}
S. Wessel, I. Niesen, J. Stapmanns, B.~Normand, F. Mila, P. Corboz, and A. Honecker, Thermodynamic properties of the Shastry-Sutherland model from quantum Monte Carlo simulations,
\href{https://doi.org/10.1103/PhysRevB.98.174432}
{Phys. Rev. B \textbf{98}, 174432 (2018)}.
%
\bibitem{DEmidio_etal2020}
J. D'Emidio, S. Wessel, and F. Mila, Reduction of the sign problem near $T=0$ in quantum Monte Carlo simulations,
\href{https://doi.org/10.1103/PhysRevB.102.064420}
{Phys. Rev. B \textbf{102}, 064420 (2020)}.
%
\bibitem{Hangleiter_etal2020}
D. Hangleiter, I. Roth, D. Nagaj, and J. Eisert, Easing the Monte
Carlo sign problem,
\href{https://doi.org/10.1126/sciadv.abb8341}
{Sci. Adv. \textbf{6} eabb8341 (2020)}.
%
\bibitem{Levy-Clark2021}
R. Levy and B. K. Clark, Mitigating the sign problem through basis rotations,
\href{https://dx.doi.org/10.1103/PhysRevLett.126.216401}
{Phys. Rev. Lett. \textbf{126}, 216401 (2021)}.
%
\bibitem{Murota-Todo2025}
K. Murota and S. Todo, Local basis transformation to mitigate negative sign problems,
\href{https://doi.org/10.48550/arXiv.2501.18069}
{arXiv:2501.18069.}


\bibitem{Mazzola2024}
G. Mazzola, Quantum computing for chemistry and physics applications from a Monte Carlo perspective,
\href{https://doi.org/10.1063/5.0173591}
{J. Chem. Phys. \textbf{160}, 010901 (2024)}.
%
\bibitem{Jiang_etal2025}
T. Jiang, J. Zhang, M. K. A. Baumgarten, M.-F. Chen, H. Q. Dinh, A. Ganeshram, N. Maskara, A. Ni, and J. Lee, Walking through Hilbert space with quantum computers,
\href{https://pubs.acs.org/doi/full/10.1021/acs.chemrev.4c00508}
{Chem. Rev. \textbf{125}, 4569 (2025)}.


\bibitem{Tan_etal2022}
K. C. Tan, D. Bhowmick, P. Sengupta, Sign-problem free quantum stochastic series expansion algorithm on a quantum computer,
\href{https://doi.org/10.1038/s41534-022-00555-x}
{npj Quantum Inf. \textbf{8}, 44 (2022)}.


\bibitem{SSEreview_1}
A. W. Sandvik, Computational Studies of Quantum Spin Systems,
\href{https://doi.org/10.1063/1.3518900}
{AIP Conf. Proc. \textbf{1297}, 135 (2010)}.
%
\bibitem{SSEreview_2}
A. W. Sandvik, Stochastic series expansion methods, \emph{Many-Body Methods for Real Materials, Modeling, and Simulation}, edited by E. Pavarini, E. Koch, and S. Zhang (Verlag des Forschungszentrum Julich, 2019), Vol. 9.


\bibitem{Hen2021}
I. Hen, Determining quantum Monte Carlo simulability with geometric phases,
\href{https://doi.org/10.1103/PhysRevResearch.3.023080}
{Phys. Rev. Research \textbf{3}, 023080 (2021)}.


\end{thebibliography}
\end{document}